\documentclass[lineno]{jfm}

\usepackage{graphicx}
\usepackage{newtxtext}
\usepackage{newtxmath}
\usepackage{natbib}
\usepackage{hyperref}
\hypersetup{
    colorlinks = true,
    urlcolor   = blue,
    citecolor  = black,
}

\newcommand{\RomanNumeralCaps}[1]
\nolinenumbers
\usepackage{subfigure}
\usepackage{xcolor}

\title{\nolinenumbers Flux scaling in Rayleigh B\'enard convection: a local boundary layer analysis}

\author{\nolinenumbers Prafulla P. Shevkar\aff{1}
  \corresp{\email{prafulla145@gmail.com}},
 \and Baburaj A. Puthenveettil\aff{1}}

\affiliation{\nolinenumbers \aff{1}Department of Applied Mechanics, Indian Institute of Technology Madras, Chennai, India 600036
}

\begin{document}
\maketitle
\begin{abstract}
\nolinenumbers
We study the effect of shear due to the large scale flow (LSF) on the heat flux in Rayleigh B\'enard convection for a range of near-plate Rayleigh numbers $8\times 10^7 \leq Ra_w \leq 5\times 10^{14}$, by studying its effect on the local boundary layers (BLs) on either sides of the plumes, which are much thinner than the global shear BL created by the LSF velocity $V_F$. Considering these local BLs forced externally by the LSF, we obtain a fifth order algebraic equation for the local boundary layer thicknesses using the order of magnitude balance of the corresponding mixed convection BL equations. Solving these equations numerically using the observed Reynolds number relations for the LSF strengths, for aspect ratios $\Gamma=1$ and 0.5, we obtain the variation of the local BL thicknesses with the longitudinal distance for various $Ra_w$. We find that the average shear acting on the edges of these local BLs increases as $\overline{u|}_{z=\delta} \sim Ra_w^{1/3}$ for $8\times 10^7\leq Ra_w \leq 10^{12}$ at $\Gamma=1$, and as $\overline{u|}_{z=\delta} \sim Ra_w^{0.38}$ for $1\times 10^{11}\leq Ra_w \leq  5\times 10^{14}$ at $\Gamma=0.5$. Correspondingly, the longitudinal development of these BLs deviate more from that in natural convection boundary layers (NCBL); the local BLs then become more of mixed convection nature with increase in $Ra_w$. We observe that the average dimensionless shear acting at the edge of these local BLs decreases with increasing $Ra_w$ as $\overline{A}=u|_{z=\delta}/V_F\sim Ra_w^{-0.102}$ for $\Gamma=1$ and $\overline{A}\sim Ra_w^{-0.089}$ for $\Gamma=0.5$. This observation implies that the average shear forcing of these local BLs ($\overline{u|}_{z=\delta}$) increases less, compared to the corresponding increase of $V_F$ with $Ra_w$. We then estimate the average local thermal BL thickness ($\overline{\delta_T}$) by spatially averaging these local BL thicknesses over the mean plume spacing in the presence of shear, to find the global Nusselt number $Nu=H/2\overline{\delta_T}$, where $H$ is the fluid layer height. We find that $Nu\sim Ra_w^m$, where $m\approx 0.327$ for $8\times 10^7 \leq Ra_w \leq 1\times 10^{12}$ at $\Gamma=1$, and $m=0.33$ for $1\times10^{11}\leq Ra_w \leq 5\times10^{14}$ at $\Gamma=0.5$. Inspite of the increasing shear on these BLs with increasing $Ra_w$, we then surprisingly obtain the classical 1/3 scaling of flux, with no transition to ultimate regime seen. We then show that the absence of any transition in the flux scaling towards an ultimate regime upto $Ra_w\leq 5\times10^{14}$, inspite of the increasing mixed convection nature of these local BLs, occurs since the shear forcing acting on those BLs remains sub-dominant compared to the NCBL velocities ($V_{bl}$) within these BLs.
\end{abstract}

\nolinenumbers

\section{\label{sec:intro}Introduction }
 Rayleigh-Bénard convection, a type of natural convection occurring between two horizontal plates maintained at different temperatures, has been extensively studied due to its complex and rich fluid dynamics. A key focus in this area is understanding how the Nusselt number $Nu$, which measures the efficiency of convective heat transfer, scales with the Rayleigh number $Ra_w$. Specifically, the scaling relationship $Nu\sim Ra_w^m$ has been a topic of contention, with different values of the exponent $m$ reported in the literature~\citep{castaing1989scaling,Roche_2010,Lohse_ultimateregime_2011,scheel_kim_white_2012,he2012_ultimateregm,Iyer2019_classical1_3scalingtill10power15}.
  
 The classical scaling exponent is 1/3 where the near-wall Rayleigh number $Ra_w=g\beta\Delta T_w H^3/\nu\alpha$. Here, $\beta$ is the coefficient of thermal expansion, $g$ the acceleration due to gravity, $\nu$ the kinematic viscosity, $\alpha$ the thermal diffusivity, $H$ the fluid layer
height and $\Delta T_w$ the
temperature difference between the hot plate temperature $T_h$ and the bulk fluid temperature $T_b$; $\Delta T=2\Delta T_w$, where $\Delta T=T_h-T_c$ with $T_c$ being cold plate temperature, then Rayleigh number $Ra=2Ra_w$. 

For $Ra_w<10^{13}$, there seems to be a broad acceptance that the scaling exponent has a value of $2/7<m<1/3$ \citep{scheel_kim_white_2012,LamXiaetc_Prdependence2002,Chao_etal2022_Nu0.33}. Also, most studies find the Reynolds number based on the large-scale flow (LSF) velocity $V_F$ scaling as $Re \sim Ra^{0.43}$ \citep{LamXiaetc_Prdependence2002,puthenveettil05:_plume_rayleig,ahlers09:_heat_rayleig_benar,he2012_ultimateregm,gunasegarane14:_dynam}, where $Re=V_FH/\nu$. With increased $Ra_w$, the strength of the LSF increases. This stronger LSF, when interacting with the near-wall region, also increases velocities within the local boundary layers (BLs) close to the hot plate from their natural convection values given by  $V_{bl}$~\citep{shevkar_mohanan_puthenveettil_2023}. The Reynolds number based on $V_{bl}$ is 
\begin{eqnarray}
\label{eq:Rebl}
Re_{bl}=V_{bl}H/\nu=1.88Ra_w^{1/3}Pr^{-0.98},
\end{eqnarray} 
\citep{puthenveettil11:_lengt}.  As $Ra_w$ increases, the thermal boundary layer thickness $\delta_T$ decreases rapidly. Consequently, the dimensionless heat transfer coefficient, $Nu$, which is proportional to $H/2\delta_T$, increases. 

At even higher values of  $Ra_w$, $10^{10} \leq Ra_w\leq 10^{15}$, \cite{Iyer2019_classical1_3scalingtill10power15} demonstrates that the flux scaling approaches the classical 1/3 scaling at aspect ratio $\Gamma=L/H=0.1$, where $L$ is the length of the convection cell. However, contradictory inferences report a transition to the so-called ultimate regime, which is observed as a scaling of $Nu\sim Ra_w^{0.38}$, at $Ra\approx 10^{14}$ \citep{he2012_ultimateregm,Lohse_ultimateregime_2011}. \cite{he2012_ultimateregm} attributed this transition in flux scaling to a change in the scaling of Reynolds number $Re$ on $Ra_w$ from $Re\sim Ra_w^{0.43}$ to $Re\sim Ra_w^{0.5}$, claiming a different boundary layer dynamics at higher values of $Ra_w$. 

The aspect ratio $\Gamma$, depending on which the large-scale dynamics significantly varies, also plays a role in deciding the flux scalings \citep{Ahlers22_AspectratioDependence}.
Recently, \cite{samuel2024boundary} studied RB convection in $\Gamma=4$ and 8 channels for $Ra_w\le10^{11}$ relevant to geophysical applications, and showcases that the BLs are fluctuation dominated with fluctuations being much higher in magnitudes that the mean flow. In the absence of persistent LSF in high aspect ratio channels, these results raise questions over the claimed shear-induced transition in flux scalings. 

Majority of flux scalings were discovered by spatially and temporally averaging thermal boundary layer thicknesses, and the transition in flux scaling was subsequently claimed using these averaged values of BL thicknesses~\citep{grossman00:_scalin,Ahlers22_AspectratioDependence}.  In reality, there are local boundary layers on the plates that turns into plumes due to gravitational instability \citep{Pera73}. This instability is modified with increasing shear effects \citep{castaing1989scaling}, resulting in an increased spacing between plumes ($\lambda_s$) from its no-shear value ($\lambda_0$) \citep{shevkar19}. The plume spacing $\lambda_s$, in the presence of small shear, normalised by the viscous-shear length $Z_{sh}=\nu /V_F$, is given by
    \begin{equation}
    \label{eq:lam_lam0}
    {\lambda_s}=\lambda_0+\frac{{Z_{sh}}Re^3}{D Ra_w}, 
    \end{equation}
    where, 
    $D(Pr)=0.004Pr^3$ for $Pr>5$ and $D(Pr)=52.7Pr^{-2.8}$ for $Pr<5$. These local BLs, which feed plumes from the sides, are significantly thinner compared to the the global BL thickness, obtained by spatio-temporally averaging the temperature within BLs and plumes. Rather than the global BL, a near-wall temperature drop ($\Delta T_w$) that occurs on thinner, local BLs present on either side of plumes are the ones that decides the flux scalings in Rayleigh-Bénard convection. Moreover, averaging the BL and plume regions together essentially considers differently characterized regions of strain rates as one~\citep{shevkar22_plumedetection,shevkar2023_tomoPIVinvesti}.


    Several models have been proposed by different authors to explain the scalings of Nusselt number in Rayleigh-Bénard convection. The formulation of models by various authors exhibits a notable contrast in their approach, as they assume distinct natures of BLs. Initially, \cite{Howard1966} considered a marginally stable BLs assuming that the LSF has no impact on the stability of the BLs close to the plates. However, this neglected the interaction between the thermal plumes and the LSF~\citep{shevkar19}. Therefore, the revisited model of Malkus and Howard recently developed by \cite{CREYSSELS202497}, considering the LSF as an external input, obtained $Nu$ values similar to those in experiments and simulations for $Ra<10^{10}$. \cite{Kraichnan1962} proposed a model with nature of the BL analogous to that in a fully developed shear flow and the ultimate scaling of 1/2 at extremely high $Ra$. \cite{castaing1989scaling} developed a model in which the domain was divided into near-wall BLs, turbulent bulk, and intermediate mixing region, and proposed a flux scaling with exponent of $Ra_w$ equal to 2/7, in agreement with the observed values in the experiments for $Ra<10^{13}$. \cite{ShraimanSiggia} proposed a model presuming the fully turbulent nature of boundary layers leading to 2/7 scaling. While supporting the transition in flux scalings at $Ra_w\sim 10^{14}$, \cite{Lohse_ultimateregime_2011} argued that the nature of the boundary layers changes from laminar at low $Ra$ to fully turbulent at extremely high $Ra$. \cite{skrbek_urban_2015} re-analysed experimental data of \cite{Roche_2010} and \cite{he2012_ultimateregm} at very high $Ra$ and attributed the so-called transition to ultimate regime to non-Oberbeck–Boussinesq effects. Recently, a theory on BLs by \cite{lindborg_2023} hypothesised the near-wall boundary layers of a semi-turbulent nature, with a presence of thin viscous wall layer of about five Kolmogorov scales at $Ra_w$ as high as $10^{15}$, and concluded that $Nu\sim Ra_w^{1/3}$ as the scaling of the ultimate regime. 
    
    Summing up the investigations in the literature, we propose a BL model which describes the near-wall dynamics in Rayleigh B\'enard convection for $8\times 10^7 \leq Ra_w \leq 5\times10^{14}$. We consider the two-dimensional laminar natural convection type BLs, which are thinner than the global BL. These are local BLs on the sides of the plumes, forced with the mean Prandtl-Blasius velocity profile generated by the LSF. The LSF velocity scalings are provided as an external input employed from the literature studies~\citep{cioni,Roche_2010}. A fifth order scaling equation for the local BLs forced by the shear due to the LSF is obtained using the order of magnitude balance of integral BL equations, which was then solved numerically. Calculating averaged values of the thermal BLT $\overline{\delta_T}$, over the half mean plume spacing ($\lambda_s/2$) \eqref{eq:lam_lam0}, the variation of the Nusselt number $Nu$ with $Ra_w$ is then computed. The scaling of $Nu$ follows the power law $Nu\sim Ra_w^{0.327}$ for $Ra_w\leq 10^{12}$ at $\Gamma=1$ while $Nu\sim Ra_w^{0.33}$ for $1\times10^{11}\leq Ra_w \leq 5\times10^{14}$ at $\Gamma=0.5$. We demonstrate the reasoning behind the classical 1/3 scaling of the Nusselt number up to $Ra_w=10^{15}$ by relating it to the relatively high strength of buoyancy effects compared to shear effects inside the BLs.
\section{Analysis of the local BLs forced by large scale flow (LSF)}
		\begin{figure}
	    \centering
\subfigure[]{\includegraphics[width=0.6\textwidth, trim=0 0 0 0,clip]{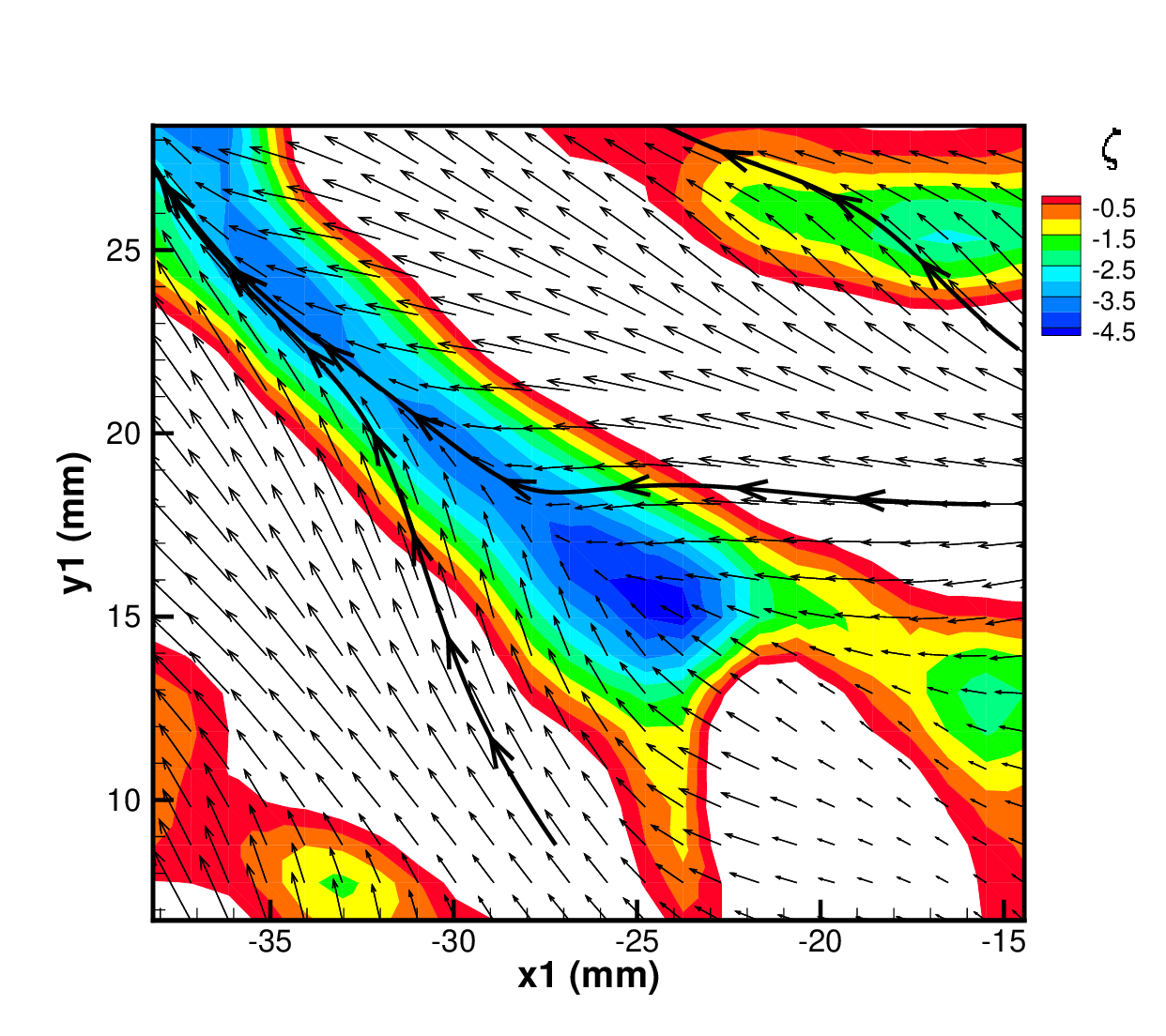}\label{fig:hd}}\\
\subfigure[]{\includegraphics[width=0.49\textwidth]{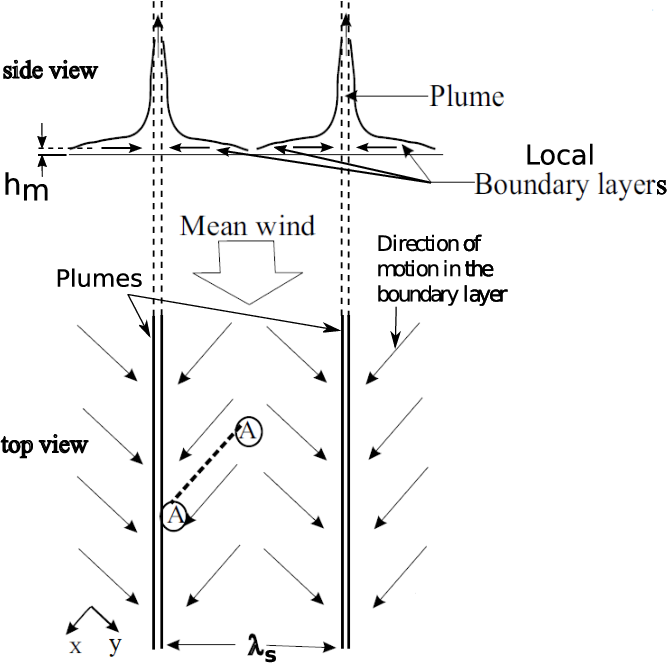}\label{fig:topview}}
		\subfigure[]{\includegraphics[width=0.49\textwidth]{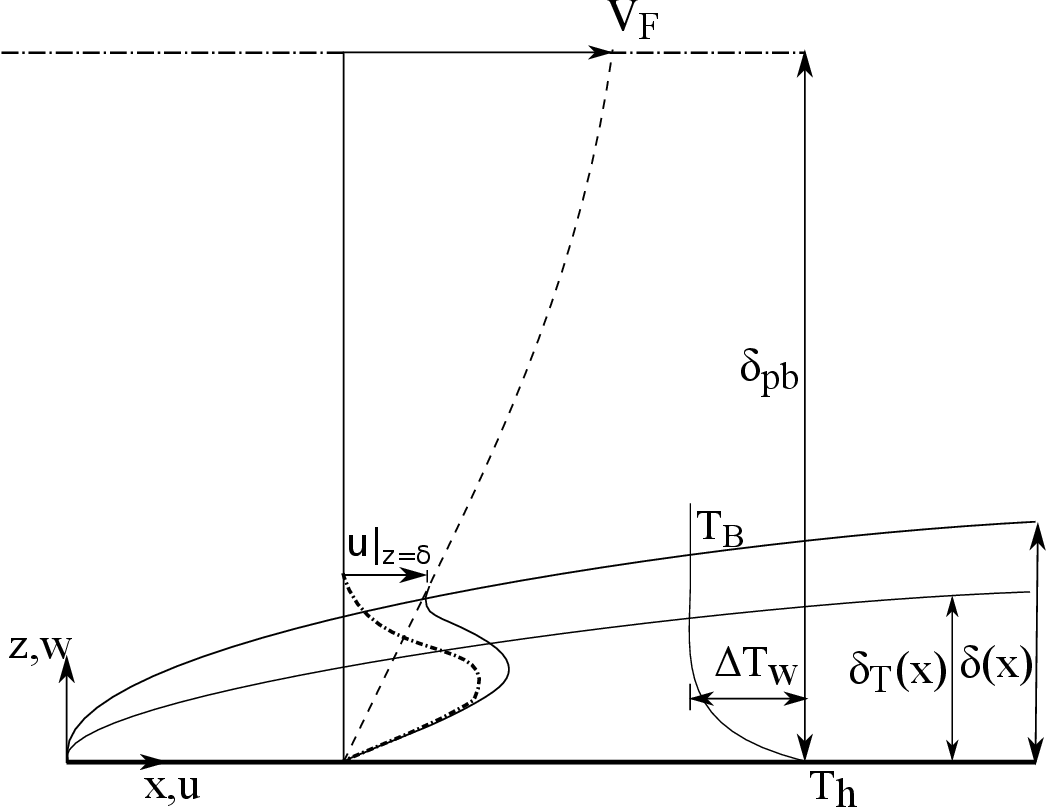}\label{fig:alongAA}}   	     
	      
	     \caption{(a)Schematic of the local boundary layers between line plumes being forced by an external shear due to the large-scale flow for $Pr>1$. The top schematic shows the side view  in a vertical plane through B--B while the bottom schematic shows the corresponding top view; (b) Top view of the horizontal velocity field obtained using PIV, overlaid over the corresponding negative horizontal divergence field, in a horizontal plane at a height of $h_m=1.6$mm from the hot plate, at $Ra_w=1.31\times10^7$ \citep{shevkar19,shevkar22_plumedetection,shevkar_mohanan_puthenveettil_2023}. The white regions show the local BLs and the colored regions show the plumes. (c) Side view in a vertical plane through A--A in figure (a); ---, horizontal velocity profile within the local BL due to forcing by the external shear of the LSF on the natural convection BL; $\color{black}-\cdot-\cdot-$, horizontal velocity profile within a NCBL; $---$, horizontal velocity profile within a PBBL.
	    }
	    \label{fig:local_bl}
	\end{figure}

Figure~\ref{fig:hd} shows the top view of the actual local flow field in a horizontal plane close to the hot plate, obtained by PIV, after applying the plume separation criterion by \cite{shevkar22_plumedetection}. The figure shows line plumes (coloured regions) aligned by an LSF oriented from bottom left corner to top left corner of the figure. The white regions in the figure are the local BLs in between the plumes, that are externally forced by the LSF. The schematic of the side view, in the vertical plane through B--B in figure \ref{fig:hd} is shown in the top figure in \ref{fig:topview}. The direction of LSF in this figure is perpendicular to the plane of the figure. The corresponding schematic of the top view, showing the plumes aligned by the LSF and the local BLs in-between those plumes. A side view of one of such local BL, in vertical plane through A--A in figure~\ref{fig:topview} is shown in figure~\ref{fig:alongAA}. These local BLs are an order thinner than the global BL created by the LSF, which span the length of the convection cell~\citep{gunasegarane14:_dynam,shevkar19,shevkar22_plumedetection}. Since the major part of the temperature drop near the plate occurs across these BLs, these thinner local BLs are more likely to be the ones which decide the flux scaling near the hot surface and not the global BLs as has been assumed so far. Due to their smaller thickness as well as since they are embedded within the global BL, these local BLs are less likely to be affected by the turbulent bulk. These local boundary layers on the hot plate become unstable to turn upwards and form the plumes. Since plumes are the outcome of the instability of these thin local BLs, on their either sides, it is natural to expect that these BLs, before they become unstable, to be laminar. Further, successful scaling laws for the mean spacing
between such line plumes \citep{theerthan98:_rayleig,theerthan00:_planf, puthenveettil05:_plume_rayleig},
their total lengths \citep{puthenveettil11:_lengt}, their mean dynamics (\citep{gunasegarane14:_dynam}) and most recently the velocities within local BLs~\citep{shevkar_mohanan_puthenveettil_2023} have all been obtained assuming  steady 2D laminar natural convection boundary layers
\citep{rotem69:_natur,Pera73_ncbl} feeding these plumes. More importantly, the observed vertical distributions of fluctuations of velocities and temperature near the hot plate have been predicted well by a model that assumes laminar BLs giving rise to laminar plumes \citep{theerthan98:_rayleig}. This means that the observed fluctuations in velocity and temperature near the hot-plate are a result of the spatial averaging of a spatially non-uniform field consisting of many local laminar BLs giving rise to many plumes with the bulk fluid in between them, as well as by the lateral motion of such plumes \citep{shevkar22_fluct}. 
Due to all these reasons we consider these local BLs to be 2D laminar natural convection boundary layers on a horizontal surface, forced by an external shear due to a LSF of strength $V_{F}$, as shown in figure~\ref{fig:local_bl}. Since these local BLs are affected by the shear due to the LSF, in other words, we assume that the local boundary layers that occurs on either sides of the plumes on the plate in turbulent convection, which we use to find the flux scaling, to be of laminar mixed convection nature. 
\subsection{Integral relations}
\label{sec:integraleqs}
The $x$-momentum BL equation, 
 $u{\partial u}/{\partial x} + w{\partial u}/{\partial z} = -({1}/{\rho}) {\partial p}/{\partial x}+\nu {\partial^2 u}/{\partial z^2}$, integrated across the velocity BL thickness $\delta(x)$ for such BLs
is,
	\begin{eqnarray}
	\label{eq:xmomen0}
	\hspace{-0.3cm}\int_0^{\delta} u\frac{\partial u}{\partial x} dz+\int_0^{\delta} w\frac{\partial u}{\partial z} dz= -\frac{1}{\rho}\int_0^\delta\frac{\partial p}{\partial x}dz+\nu \int_0^{\delta} \frac{\partial^2 u}{\partial z^2} dz ,
	\end{eqnarray}
	where $p(x,z)$ is the pressure within the boundary layer and $u$ $\&$ $w$ the horizontal and vertical velocity components, respectively.
	Rewriting the second term in (\ref{eq:xmomen0}) as $\int_0^{\delta} \left({\partial(wu)}/{\partial z} - u{\partial w}/{\partial z}\right)dz$, integrating, replacing ${\partial w}/{\partial z}=-{\partial u}/{\partial x}$ from continuity, noticing that $\int_0^{\delta} \left(u{\partial u}/{\partial x}\right) dz =\int_0^{\delta} \left({\partial u^2/\partial x}\right) dz/2$, and since ${\partial u/\partial z}|_{z=\delta}\simeq~0$ for small external shear, we rewrite \eqref{eq:xmomen0} as
	\begin{eqnarray}
	\label{eq:xmomen1}
	\int_0^{\delta} \frac{\partial}{\partial x}u^2 dz+\left.(w u)\right|_{z=\delta}=\frac{-1}{\rho} \int_0^{\delta}\frac{\partial p}{\partial x}dz-\nu\left.\frac{\partial u}{\partial z}\right|_{z=0}.
	\end{eqnarray}
 
    In the second term in (\ref{eq:xmomen1}), the vertical velocity at the velocity boundary layer edge can be obtained by integrating the continuity equation, $\partial u/\partial x+\partial w/\partial z=0$, and applying the Leibnitz rule as
	\begin{equation}
	\label{eq:w}
	w|_{z=\delta}=-\frac{\partial }{\partial x}\int_0^\delta udz+ \left.u\right|_{z=\delta} \frac{\partial \delta}{\partial x}.
	\end{equation}
	To obtain the term $\left.u\right|_{z=\delta}$ which is the external forcing of these local BLs, that occur in (\ref{eq:xmomen1}) and (\ref{eq:w}), we assume that these local boundary layers are embedded within a PBBL spanning the entire length of the hot plate and which is driven by the large-scale flow strength of $V_{F}$, acting at a distance of $\delta_{pb}$ from the hot plate; the schematic of such an arrangement is shown in figure~\ref{fig:alongAA}. Using the Von-Karman velocity profile 
 \begin{equation}
     \label{eq:von_karmann}
      {u(z)}/V_F=2{z}/{\delta_{pb}}-\left({z}/{\delta_{pb}}\right)^2,
 \end{equation}
 for the PBBL, the dimensionless shear velocity acting on the edges of the local natural convection boundary layers, at a height of $z=\delta$ from the hot plate, is then,
	\begin{equation}
	    \label{eq:A}
	    A(x)=\frac{\left.u\right|_{z=\delta} }{V_{F}}=2\frac{\delta}{\delta_{pb}}-\left(\frac{\delta}{\delta_{pb}}\right)^2,
	\end{equation}
	where the mean PBBL thickness, independent of $x$, is 
    \begin{equation}
        \label{eq:PBBL}
        \delta_{pb}=0.922H/\sqrt{Re}
    \end{equation}
 \citep{ahlers09:_heat_rayleig_benar,stevens_etc_prefactor_2013}. 
	
 Applying Leibnitz rule, to the first term in \eqref{eq:xmomen1}, substituting \eqref{eq:w} in \eqref{eq:xmomen1}, simplifying, and replacing $\left.u\right|_{z=\delta}$ using \eqref{eq:A}, we obtain,
	\begin{eqnarray}
	\label{eq:xmomen}
	\frac{\partial}{\partial x} \int_0^{\delta} u^2dz-A(x) V_{F}\frac{\partial}{\partial x}\int_0^\delta udz+\frac{1}{\rho}\int_0^\delta\frac{\partial p}{\partial x}dz+\nu \left.\frac{\partial u}{\partial z}\right|_{z=0}=0.
	\end{eqnarray}
    To replace the unknown pressure in (\ref{eq:xmomen}), we integrate the z-momentum BL equation, ${\partial p/\partial z}=\rho g \beta(T-T_B)$, across the velocity BL thickness, to obtain,	
    \begin{equation}
	\label{eq:ymomen}
	p=-\rho g \beta \int_0^\delta(T-T_B)dz,
	\end{equation}
	where $T(z)$ is the temperature distribution within the local BL, and $T_B$ the fluid temperature above the local BLs. Substituting (\ref{eq:ymomen}) in (\ref{eq:xmomen}), we obtain the integral momentum balance equation for the local natural convection boundary layers, forced externally by the large-scale flow, as
	\begin{eqnarray}
	\label{eq:xmomen2}
	\hspace{-0.1cm}\frac{\partial}{\partial x} \int_0^{\delta} &&u^2dz-A V_{F}\frac{\partial}{\partial x}\int_0^\delta udz-g\beta\int_0^\delta\frac{\partial }{\partial x} \int_0^\delta(T-T_B)dz dz+\nu \left.\frac{\partial u}{\partial z}\right|_{z=0}=0.
	\end{eqnarray}
		
	The local BL energy equation, 
	$ u{\partial T}/{\partial x}  + w{\partial T}/{\partial z} = \alpha {\partial^2 T}/{\partial z^2 }$, integrated across the local thermal BL thickness $\delta_{T}(x)$, is
	\begin{equation}
	\label{eq:energy0}
	\int_0^{\delta_T} u\frac{\partial T}{\partial x} dz +\int_0^{\delta_T} w\frac{\partial T}{\partial z} dz= \alpha \int_0^{\delta_T} \frac{\partial^2 T}{\partial z^2 }dz.
	\end{equation}
    The first term in (\ref{eq:energy0}) can be written as $\int_0^{\delta_T} \left({\partial(uT)}/{\partial x}\right) dz-\int_0^{\delta_T} T\left({\partial u}/{\partial x}\right) dz$. Applying Leibnitz rule to the first term in this equation, the first term in (\ref{eq:energy0}) becomes, 
	\begin{eqnarray}
	\label{eq:hash1}
	\frac{\partial}{\partial x}\int_0^{\delta_T} uT dz-\left.(uT)\right|_{z=\delta_T}\frac{d \delta_T}{dx}-\int_0^{\delta_T} T\frac{\partial u}{\partial x}dz.
	\end{eqnarray}
	Similarly, we rewrite the second term in (\ref{eq:energy0}) as $\int_0^{\delta_T} \left({\partial(wT)}/{\partial z}\right) dz-\int_0^{\delta_T} T\left({\partial w}/{\partial z}\right) dz$. Integrating the first term of this equation, and using continuity equation on the second term of this equation, the second term in (\ref{eq:energy0}) becomes, 
	\begin{eqnarray}
	\label{eq:hash2}
	\left.(wT)\right|_{z=\delta_T}+\int_0^{\delta_T}T\frac{\partial u}{\partial x}dz. 
	\end{eqnarray}
	Replacing $w|_{z=\delta_T}$ in (\ref{eq:hash2}) by $-{\partial}/{\partial x}\int_0^{\delta_T}udz+u|_{z=\delta_T}\left({d\delta_T}/{dx}\right)$, obtained by integrating the continuity equation across the thermal BL thickness, similar to (\ref{eq:w}), and applying Leibnitz rule, the second term in (\ref{eq:energy0}) becomes, 
	\begin{eqnarray}
	\label{eq:hash3}
	-\frac{\partial}{\partial x}\int_0^{\delta_T}uT_Bdz+u|_{z=\delta_T}T_B\frac{d\delta_T}{dx}+\int_0^{\delta_T}T\frac{\partial u}{\partial x}. 
		\end{eqnarray}
	Substituting (\ref{eq:hash1}) and (\ref{eq:hash3}) in (\ref{eq:energy0}) and simplifying, since the last term in (\ref{eq:energy0}) becomes $-\alpha {\partial T}/{\partial z}|_{z=0}$ since $ {\partial T}/{\partial z}|_{z=\delta}\simeq0$, we obtain the integral energy equation for the local NCBLs forced by the LSF as
	\begin{equation}
	\label{eq:energy}
	\frac{\partial}{\partial x} \int_0^{\delta_T} u(T-T_B)dz=-\alpha \left. \frac{\partial T}{\partial z} \right|_{z=0}.
	\end{equation}
 \subsection{Scaling relation for the local BL thickness}
 \label{sec:scaling rels}
	We now convert the integral equations (\ref{eq:xmomen2}) and (\ref{eq:energy}) into scaling relations, using the following relevant characteristic scales near the hot/cold plate in turbulent convection.
   We take the characteristic scale of temperature difference as $\Delta T_w$, and the local characteristic vertical distances within the velocity BL and the thermal BL as the velocity and thermal BL thicknesses, $\delta(x)$ and $\delta_T(x)$, respectively. The characteristic horizontal velocity within the BLs, when the BLs are forced externally by the shear due to the LSF, is not known apriori; we take it as $U_c$, whose value we find later. Substituting these characteristic scales in the order of magnitude balances of \eqref{eq:xmomen2} and \eqref{eq:energy}, we obtain, respectively,
	\begin{equation}
	\label{eq:scal_mom}
	\frac{U_c^2 \delta}{x}-\frac{A(x) V_{F}U_c\delta}{x}-\frac{g\beta \Delta T_w \delta^2}{x}+\frac{\nu U_c}{\delta}\sim0, \;\;\mathrm{and}\;\;
	\end{equation}
	\begin{equation}
	\label{eq:scal_energy}
	\frac{U_c\Delta T_w \delta_T}{x}\sim\frac{\alpha \Delta T_w}{\delta_T}.
	\end{equation}
	For small shear forcing of the local BLs, given by $A<1$, which we show later to be the case for $Ra_w<10^{15}$, we now assume,
	\begin{equation}
	\label{eq:n_exponent}
	\frac{\delta}{\delta_T}=C_2 Pr^{n},
	\end{equation}
 	where, $C_2$ and $n$ are positive constants. Equation \eqref{eq:n_exponent} implies that we assume $\delta$ and $\delta_T$ to have the same dependence on $Ra_w$ and $Re$ in the presence of small shear forcing, so that their ratio becomes independent of $Ra_w$ and $Re$. Since $C_2$ and $n$ are positive, $\delta_T<\delta$ for $Pr>1$, with the reduction in $\delta_T$ with respect to $\delta$ being a function of $Pr$. This reasonable assumption has given a good prediction for $U_c$ in the presence of shear, earlier in \cite{shevkar_mohanan_puthenveettil_2023}.

  	Replacing $\delta_T$ in (\ref{eq:scal_energy}) with $\delta$ from (\ref{eq:n_exponent}), we obtain the expression for $U_c$ as
   \begin{equation}
       \label{eq:intermediate}
       U_c \sim \frac{\alpha x}{\delta^2}C_2^2 Pr^{2n}.
   \end{equation}
   Substituting \eqref{eq:intermediate} in (\ref{eq:scal_mom}), we obtain a scaling relation for the local velocity BL thickness $\delta(x)$, when these BLs are forced externally by the shear due to the LSF, as
	\begin{equation}
	\label{eq:del_lamn}
	\frac{Ra_{x}}{C_2^2} {\left(\frac{\delta}{x}\right)}^5+A(x)Pr^{2n}Re_{x} {\left(\frac{\delta}{x}\right)}^2 -{C_2^{2}Pr^{5n}E}\sim0,
	\end{equation}
where, $Ra_{x}=g\beta \Delta T_w x^3/\nu \alpha$ is the local Rayleigh number based on $x$. $A(x)$ is defined in \eqref{eq:A}, the dimensionless shear forcing at the edge of the local BLs due to LSF, $Re_{x}=V_Fx/\nu$ the local shear Reynolds number based on $x$, and $E(Pr)=({C_2^2Pr^{2n-1}+1})/{C_2^2Pr^{3n}}$.

\section{Results and discussion}
\subsection{Local boundary layer thicknesses and dimensionless shear}
\label{sec:LBL and dim shear}
Rewriting \eqref{eq:del_lamn} in terms of the external shear Reynolds number $(Re=V_FH/\nu)$ and the near-plate Rayleigh number ($Ra_w$), and rearranging, we obtain,
\begin{equation}
    \label{eq:del_H}
    	\frac{Ra_w}{C_2^2} \left(\frac{\delta}{H}\right)^5+APr^{2n}{Re}\frac{x}{H} \left( \frac{\delta}{H} \right)^2 -{C_2^{2}Pr^{5n}E}\left(\frac{x}{H}\right)^2\sim0.
\end{equation}
Equation \eqref{eq:del_H} describes the variation of the local BLT $\delta(x)$ with the longitudinal distance $x$ for a given $Re$ and $Ra_W$. This $\delta(x)$, in turn, decides the dimensionless external shear $A(x)$ through \eqref{eq:A}, for the appropriate dependence of $Re$ on $Ra_w$, since $\delta_{pb}$ in \eqref{eq:A} is a function of $Re$ as given by \eqref{eq:PBBL}. The expression of $Re$ varies based again on the aspect ratio $\Gamma$, as given later in \eqref{eq:cioni_Re} and \eqref{eq:Roche_Re}. The above equation has the appropriate behaviour in the limiting cases. When $Re\rightarrow0$, \eqref{eq:del_H} shows that $\delta/x\sim Ra_{x}^{-1/5}$, the expected variation in NCBL \citep{Rotem69}. When $Ra_w\rightarrow0$, $\delta/x\sim Re_{\delta_x}^{-0.5}$, where $Re_{\delta_x}=u|_{z=\delta} x/\nu$, the expected variation in PBBL. We now numerically solve \eqref{eq:del_H} and \eqref{eq:A} simultaneously at $Pr=1$ to obtain the dimensionless velocity BL thickness $\delta(x)/H$ and the dimensionless external shear $A(x)$~\eqref{eq:A}, as a function of $x$, using two Reynolds number relations proposed for two different $\Gamma$, for the corresponding ranges of near-plate Rayleigh numbers. We use
\begin{equation}
\label{eq:cioni_Re}
  Re=1.345 Ra_w^{3/7}Pr^{-0.76},
\end{equation}
for $8\times 10^7\leq Ra_w\leq 1\times 10^{12}$ at $\Gamma=1$~\citep{cioni}, and
\begin{equation}
\label{eq:Roche_Re}
  Re=0.18 Ra_w^{0.48}Pr^{-0.75},
\end{equation} 
for $1\times 10^{11}\leq Ra_w\leq 5\times 10^{14}$ at $\Gamma=0.5$ \citep{Roche_2010}. 
We take only the real and finite solutions of \eqref{eq:del_H} and \eqref{eq:A} in the present analysis.

  \begin{figure*}
		   \subfigure[]{\includegraphics[width=0.48\textwidth,trim={0.4cm 0 1.3cm 0},clip]{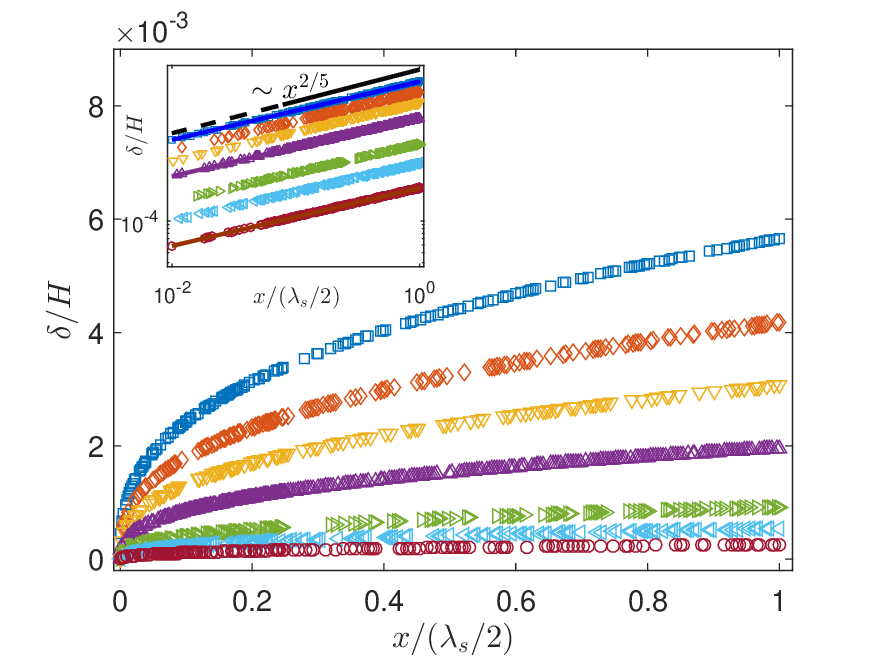}\label{fig:low_delx}}
		\subfigure[]{\includegraphics[width=0.48\textwidth,trim={0.4cm 0 1.3cm 0},clip]{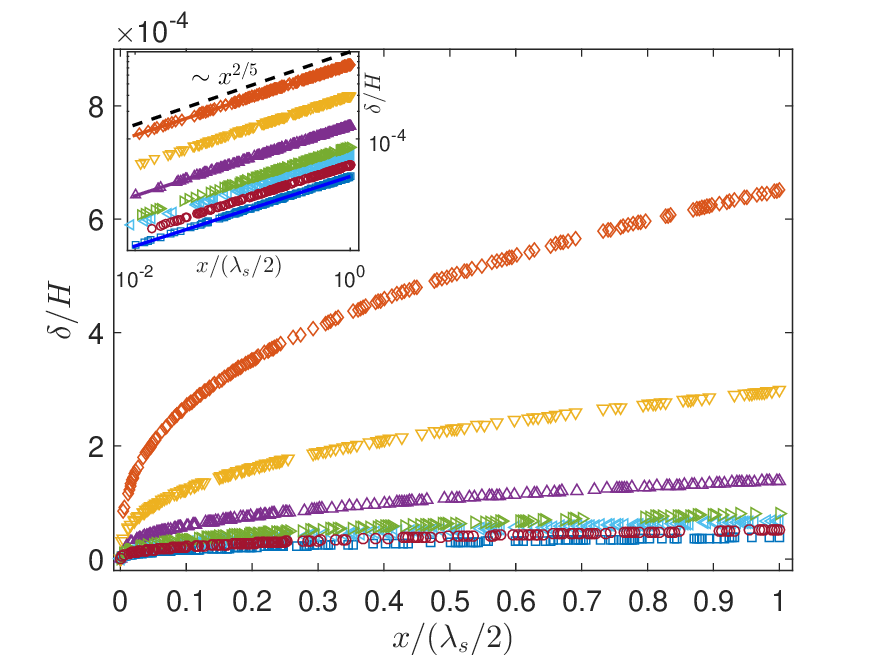}\label{fig:high_delx}}
\caption{Variation of $\delta/H$ with $x^{*}=x/(\lambda_s/2)$ for fourteen different $Ra_w$. (a), variation of $\delta/H$ obtained using \eqref{eq:cioni_Re}; $\color{blue}\square$, $Ra_w=8\times10^7$; $\color{orange}\Diamond$, $Ra_w=2\times10^8$; $\color{yellow}\triangledown$, $Ra_w=5\times10^9$; $\color{violet}\triangle$, $Ra_w=2\times10^9$; $\color{green} \rhd$, $Ra_w=2\times10^{10}$;$\color{cyan}\lhd$, $Ra_w=1\times10^{11}$ and $\color{purple}\circ$, $Ra_w=1\times10^{12}$. Log-log plot of figure~(a) is plotted as an inset. Inset of figure (a): {\color{black}$---$}, $\sim x^{*^{2/5}}$; {\color{blue}---}, $\sim {x}^{0.366}$; {\color{violet}---}, $\sim x^{*^{0.365}}$; {\color{purple}---}, $\sim x^{*^{0.364}}$. (b), variation of $\delta/H$ obtained using \eqref{eq:Roche_Re};  $\color{orange}\Diamond$, $Ra_w=1\times10^{11}$; $\color{yellow}\triangledown$, $Ra_w=1\times10^{12}$; $\color{violet}\triangle$, $Ra_w=1\times10^{13}$; $\color{green} \rhd$, $Ra_w=5\times10^{13}$;$\color{cyan}\lhd$, $Ra_w=9\times10^{13}$; $\color{purple}\circ$, $Ra_w=2\times10^{14}$ and $\color{blue}\square$, $Ra_w=5\times10^{14}$. Log-log plot of figure~(b) is plotted as an inset. Inset of figure (b): {\color{black}$---$}, $\sim x^{*^{2/5}}$; {\color{orange}---}, $\sim x^{*^{0.3875}}$; {\color{violet}---}, $\sim x^{*^{0.383}}$;  {\color{blue}---}, $\sim x^{*^{0.3803}}$.}
\label{fig:del_A_lowRaws}
\end{figure*}
  \begin{figure}
  \subfigure[]{\includegraphics[width=0.48\textwidth,trim={0.18cm 0 0.9cm 0},clip]{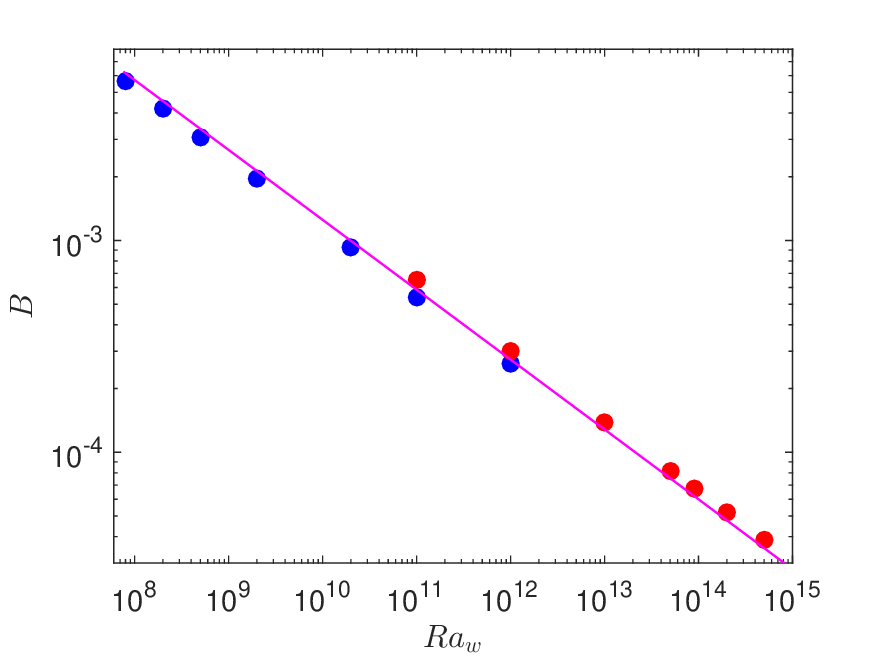}\label{fig:prefactor}}
		\subfigure[]{\includegraphics[width=0.48\textwidth,trim={0.25cm 0 0.9cm 0},clip]{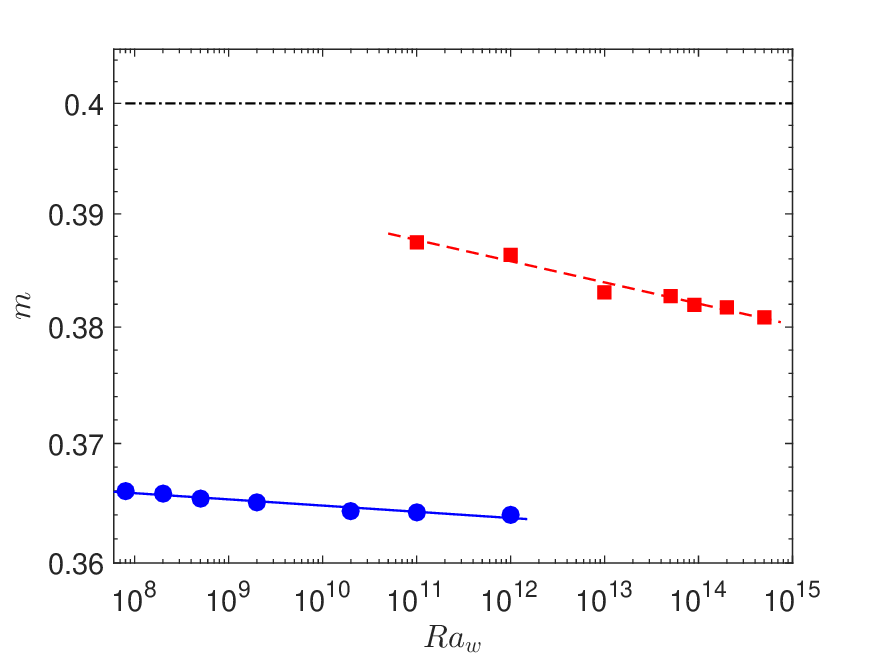}\label{fig:exponents}}
  

\caption{ Variation of (a) prefactors in the relation $\delta/H=Bx^{*^m}$ with $Ra_w$ for the data shown in figure~\ref{fig:del_A_lowRaws}. {\color{magenta}---}, $2.5 Ra_w^{-0.33}$; (b) Variation of exponents $m$ in the same; ${\color{blue}---}$, $0.37 Ra_w^{-0.00062}$; {\color{red}$---$}, $0.409 Ra_w^{-0.00211}$; $-\cdot-\cdot-$, $m=0.4$ for NCBLs.}
\label{fig:prefactor_expo}
\end{figure}
 \begin{figure*}
		\subfigure[]{\includegraphics[width=0.48\textwidth,trim={0.4cm 0 1.3cm 0},clip]{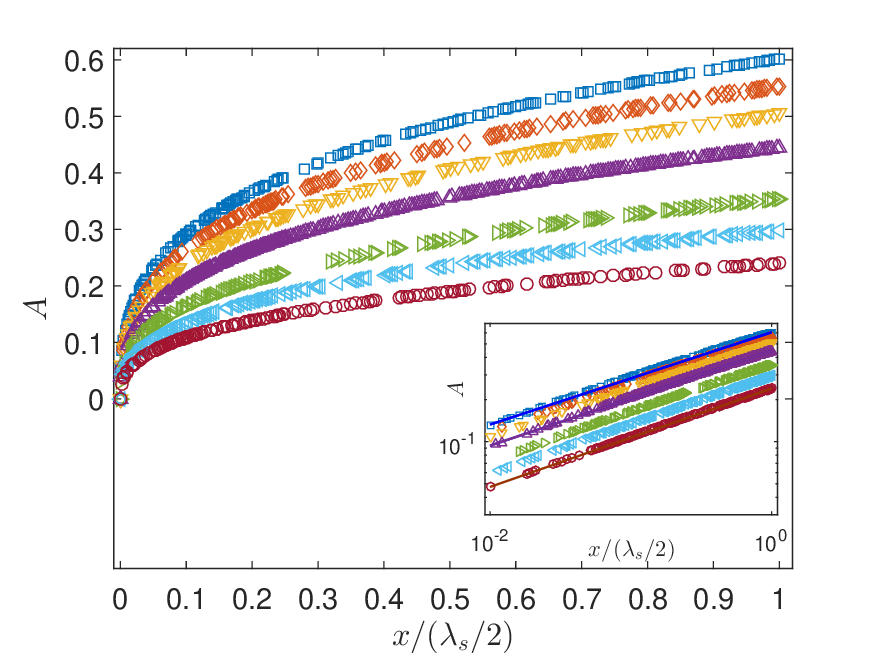}\label{fig:low_Ax}}\hfill
		\subfigure[]{\includegraphics[width=0.495\textwidth,trim={0 0 1.3cm 0},clip]{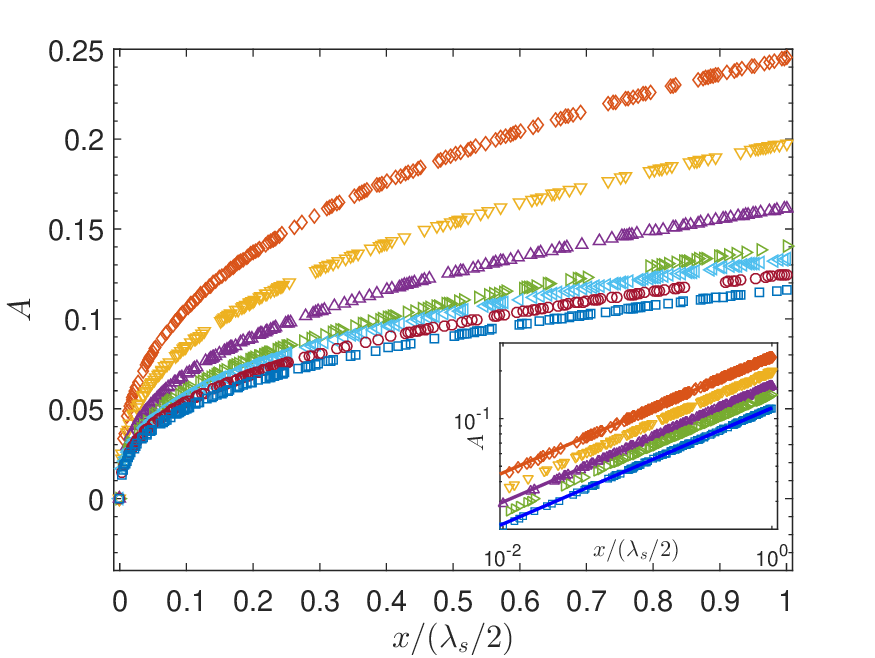}\label{fig:high_Ax}}

\caption{Variation of $A(x^{*})$ with $x^{*}$ for fourteen different $Ra_w$. In figure (a); $\color{blue}\square$, $Ra_w=8\times10^7$; $\color{orange}\Diamond$, $Ra_w=2\times10^8$; $\color{yellow}\triangledown$, $Ra_w=5\times10^9$; $\color{violet}\triangle$, $Ra_w=2\times10^9$; $\color{green} \rhd$, $Ra_w=2\times10^{10}$;$\color{cyan}\lhd$, $Ra_w=1\times10^{11}$ and $\color{purple}\circ$, $Ra_w=1\times10^{12}$. Log-log plot of figure~(a) is plotted as an inset. Inset of figure (a): {\color{blue}---}, $\sim x^{*^{0.33}}$; {\color{violet}---}, $\sim x^{*^{0.339}}$; {\color{purple}---}, $\sim x^{*^{0.352}}$. In figure (b);  $\color{orange}\Diamond$, $Ra_w=1\times10^{11}$; $\color{yellow}\triangledown$, $Ra_w=1\times10^{12}$; $\color{violet}\triangle$, $Ra_w=1\times10^{13}$; $\color{green} \rhd$, $Ra_w=5\times10^{13}$;$\color{cyan}\lhd$, $Ra_w=9\times10^{13}$; $\color{purple}\circ$, $Ra_w=2\times10^{14}$ and $\color{blue}\square$, $Ra_w=5\times10^{14}$. Log-log plot of figure~(b) is plotted as an inset. Inset of figure (b): {\color{orange}---}, $\sim x^{*^{0.3668}}$; {\color{violet}---}, $\sim x^{*^{0.3665}}$; {\color{blue}---}, $\sim x^{*^{0.3679}}$.}
\label{fig:A_lowRaws}
\end{figure*}

Figures~\ref{fig:low_delx} shows the variations of the dimensionless local BLT $\delta/H$ as a function of the dimensionless horizontal location $x^{*}=x/(\lambda_s/2)$, where $\lambda_s/2$ is half the plume spacing with shear, as obtained by using \eqref{eq:cioni_Re} for different $Ra_w$ in \eqref{eq:lam_lam0}; figure~\ref{fig:high_delx} shows the same, obtained by using \eqref{eq:Roche_Re}. The insets in figures~\ref{fig:low_delx} and ~\ref{fig:high_delx} show the same plots as in the corresponding main figures in log-log scale. Power law fits of the form $\delta/H=Bx^{*^m}$ for these $Ra_w$ are also shown in each of the inset figures. At any $x^{*}$, with increase in $Ra_w$ the values of $\delta$ decrease. Since the variation of $m$ with $Ra_w$ is small (see figure~\ref{fig:exponents}), this decrease can be quantified by the variation of the prefactors $B$, which is shown in figure~\ref{fig:prefactor}. As shown in figure~\ref{fig:exponents}, at any $x^{*}$, $\delta/H$ decreases with $Ra_w$ as $Ra_w^{-1/3}$, in the same way as the behaviour in NCBLs the absence of shear (see \cite{puthenveettil11:_lengt}). It then appears that the effect of shear does not seem to be enough to offset the decrease in $\delta$ with increase in $Ra_w$ seen in NCBLs. 

At any $Ra_w$, $\delta/H$ increase with increase in the longitudinal distance along the BL. The inset of figure~\ref{fig:low_delx} shows that for $Ra_w = 8\times 10^{7}$, $\delta/H$ approximately scale as $x^{*^{0.366}}$, which changes to $\delta/H\sim x^{*^{0.364}}$ at $Ra_w=1\times 10^{12}$. Similarly, in the inset of figure~\ref{fig:high_delx}, $\delta/H\sim x^{*^{0.3875}}$at $Ra_w=1\times10^{11}$ which changes to $\delta/H\sim x^{*^{0.3803}}$ at $Ra_w=5\times 10^{14}$. The variation of the exponents ($m$) with $Ra_w$, along with the corresponding exponent of $2/5$ for NCBL, is shown in figure~\ref{fig:exponents}. The corresponding scaling in NCBLs, namely, $\delta/H \sim x^{*^{2/5}}$, is also shown by black dashed lines in the insets of figures~\ref{fig:low_delx} and \ref{fig:high_delx}. Figure~\ref{fig:exponents} shows that, with increase in $Ra_w$, $m$ decreases in both the ranges of $Ra_w$, deviating more from the value of $m=2/5$ in NCBLs. The change in the longitudinal dependence of $\delta/H$ from that in NCBLs, imply that the shear due to the LSF changed the nature of the local BLs from a natural convection type to a mixed convection type. Then, with increase in $Ra_w$, the local BLs become more and more affected by the shear of the LSF, to become more and more of the nature of mixed convection type. As shown in figure~\ref{fig:exponents} for $\Gamma=0.5$ case, where $Re$ is given by \eqref{eq:Roche_Re}, the values of $m$ take a jump to around 0.385, which is closer than the values of $m$ for $\Gamma=1$ to the value of $m=2/5$ in NCBLs. We expect these larger values of $m$ for $\Gamma=0.5$, which implies a more natural convection nature of the local BLs, is due to the weaker LSF strengths at this lower $\Gamma$. We show later that the actual shear forcing of these local BLs, namely $u|_{z=\delta}$, that we obtain from \eqref{eq:cioni_Re} and \eqref{eq:A} is indeed lower when $Re$ is given by \eqref{eq:Roche_Re} than \eqref{eq:cioni_Re}. Further, eventhough the strengths of the LSF increase with $Ra_w$ (see \ref{eq:cioni_Re} and \ref{eq:Roche_Re}), at any $Ra_w$, their values also decrease to zero as we approach the plates (see \ref{eq:von_karmann} and figure \ref{fig:alongAA}). In such a vertically varying LSF velocity field close to the plates, a decreasing thickness of the local BLs with $Ra_w$, as could be seen to occur in figures~\ref{fig:low_delx} and \ref{fig:high_delx}, may even reduce the actual shear forcing of these local BLs with increase in $Ra_w$. To understand these issues, we now look at the variation of the actual shear forcing at the edge of these local BLs with $Ra_w$.  
		 
Figures~\ref{fig:low_Ax}~and~\ref{fig:high_Ax} show the variation of $A(x^{*})$~\eqref{eq:A}, the dimensionless shear at the edge of the BLs, with the dimensionless horizontal  location $x^{*}$, for different $Ra_w$; figure~\ref{fig:low_Ax} shows the case of $\Gamma=1$ using \eqref{eq:cioni_Re} while figure \ref{fig:high_Ax} shows the case of $\Gamma=0.5$ using \eqref{eq:Roche_Re}. For any $Ra_w$, the dimensionless shear increases with increase in $x^{*}$. The dependence of $A(x^{*})$ on $x^{*}$ for different $Ra_w$ can be better understood from the inset plots in figures \ref{fig:low_Ax} and \ref{fig:high_Ax}, which show the same plots as in the main figures in log-log scale. The inset in figure~\ref{fig:low_Ax} shows that, at $\Gamma=1$, $A(x^{*})\sim x^{*^{0.33}}$ at $Ra_w=8\times 10^7$ which changes to $A\sim x^{*^{0.352}}$ at $Ra_w=1\times10^{12}$. For $\Gamma=0.5$ case, the inset of figure~\ref{fig:high_Ax} shows that the longitudinal dependence of $A(x^{*})$ remains around $x^{*^{0.367}}$, with marginal increase to $x^{*0.368}$ at the higher $Ra_w$. Then, with increasing $Ra_w$, the exponent of $x^*$ in the longitudinal dependence of $A(x^{*})$ increases, this increase itself reduces with increasing $Ra_w$ to become marginal at the highest $Ra_w$. 

\begin{figure}
    \centering
    \includegraphics[width=0.5\linewidth, trim=0.7cm 0 01cm 0.5cm,clip]{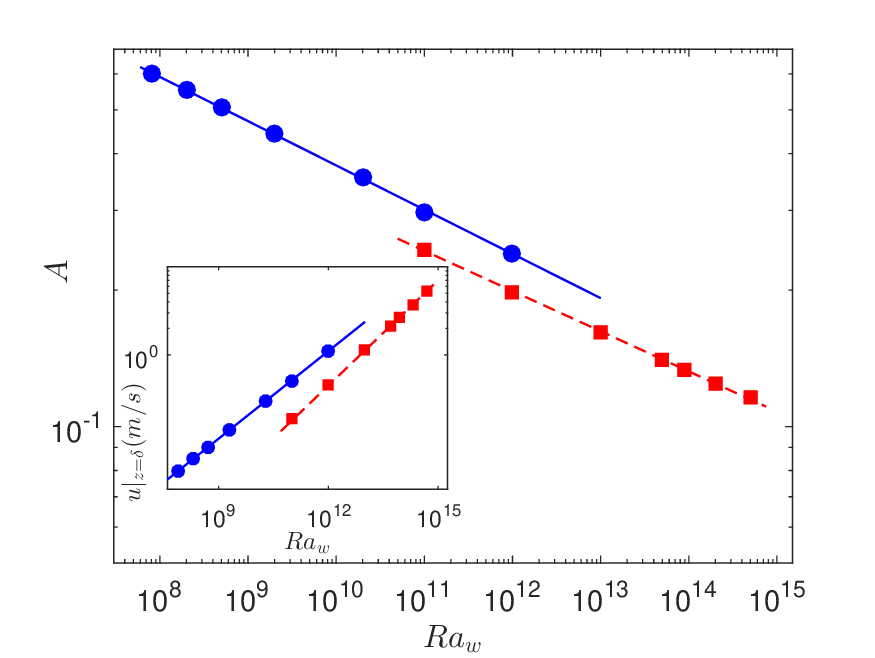}
    \caption{Variation of the local dimensionless shear at the edge of the BL with $Ra_w$; {\color{blue}---}, $\sim Ra_w^{-0.098}$ and {\color{red}$---$}, $\sim Ra_w^{-0.089}$. Inset shows the variation of the local shear velocity at the edge of BL with $Ra_w$; {\color{blue}---}, $\sim Ra_w^{0.331}$ and {\color{red}$---$}, $\sim Ra_w^{0.398}$.}
    \label{fig:Ax_Raw}
\end{figure}

Interestingly, at any $x^*$, $A(x^{*})$ decreases with increase in $Ra_w$. By plotting the prefactors of the approximately parallel lines in the insets of figures~\ref{fig:low_Ax} and \ref{fig:high_Ax}, we can find this dependence of $A(x^{*})$ on $Ra_w$ at any $x^*$, which is shown in figure~\ref{fig:Ax_Raw}. Here, the dependence of $A(x^{*})$ on $x^*$ is considered to remain approximately the same at the different $Ra_w$, since even when $Ra_w$ changes by about 6 orders, the exponent of $x^*$ changes only by $6\%$, as could be seen from figure~\ref{fig:exponents}. Figure~\ref{fig:Ax_Raw} shows that the dimensionless shear $A(x^{*})$, at any horizontal location decreases with increase in $Ra_w$ as $A(x^{*}) \sim Ra_w^{-0.098}$ for $Ra_w\leq 10^{12}$ at $\Gamma=1$ and $A(x^{*}) \sim Ra_w^{-0.089}$ for $Ra_w\geq 1\times10^{11}$ at $\Gamma=0.5$. However, at the same time,the external LSF velocities $V_F$ increase with increase in $Ra_w$ as, $V_F\sim (\nu/H)Ra_w^{3/7}$ \eqref{eq:cioni_Re} for $Ra_w\leq 10^{12}$ at $\Gamma=1$, and as $V_F\sim (\nu/H)Ra_w^{0.48}$ \eqref{eq:Roche_Re} for $Ra_w\geq1\times10^{12}$ at $\Gamma=0.5$. Since $A(x^{*})=u|_{z=\delta}/V_F$ from \eqref{eq:A}, $u|_{z=\delta}$ then scales as $(\nu/H)Ra_w^{0.331}$ for $Ra_w\leq 10^{12}$ at $\Gamma=1$, and as $(\nu/H)Ra_w^{0.398}$ for $Ra_w \geq 1\times10^{12}$ at $\Gamma=0.5$, as shown in the inset of figure~\ref{fig:Ax_Raw}. Hence, at any $x^*$, eventhough the magnitudes of the dimensionless shear (A) acting on the local BLs ($u|_{z=\delta}$) decrease with increase in $Ra_w$, the magnitudes of the actual dimensional shear acting at the edge of these local BLs do increase with increase in $Ra_w$. In agreement with this increase in $u|_{z=\delta}$ with $Ra_w$, as we saw earlier in figures~\ref{fig:low_delx}, \ref{fig:high_delx} and \ref{fig:exponents}, the longitudinal dependence of the  dimensionless local BLT $\delta(x)/H$ deviates away from the corresponding NCBL scaling, with increase in $Ra_w$, implying a shift in the nature of these BLs towards mixed convection.

   \begin{figure}
  \subfigure[]{\includegraphics[width=0.48\textwidth,trim={0.3cm 0 1cm 0},clip]{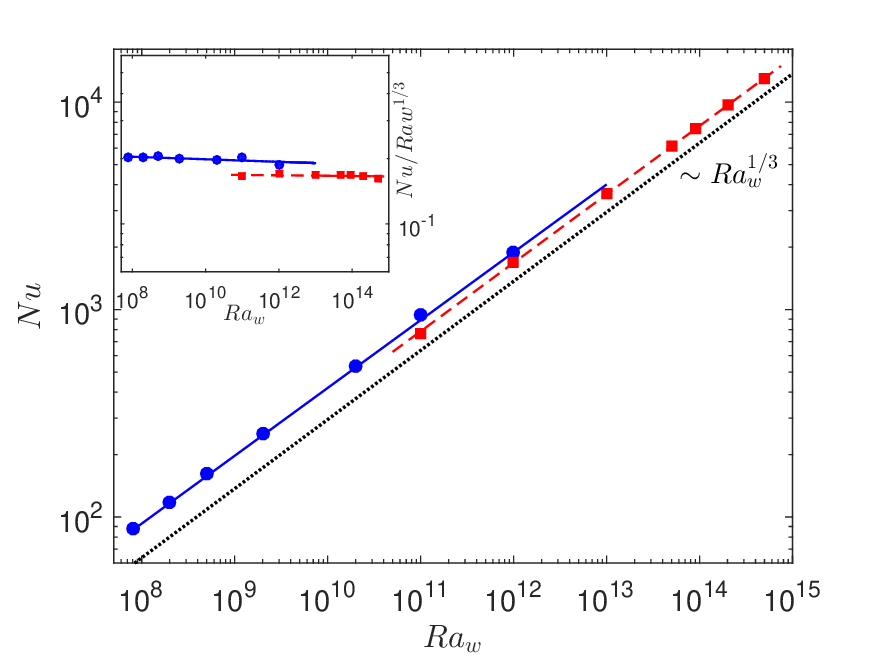}\label{fig:Nu}}
		\subfigure[]{\includegraphics[width=0.48\textwidth,trim={0.3cm 0 1cm 0},clip]{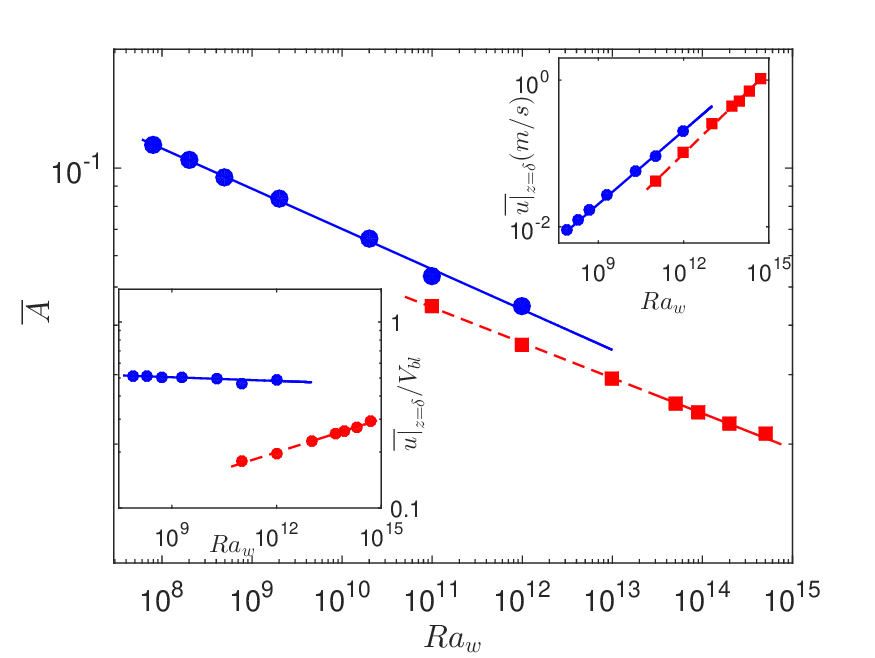}\label{fig:A_Raw}}
  

\caption{ (a) Variation of $Nu$ with $Ra_w$; {\color{blue}---}, $0.1842 Ra_w^{0.327}$ and {\color{red}$---$}, $0.225 Ra_w^{0.33}$. {\color{black}$\cdot\cdot\cdot$}, $0.137Ra_w^{1/3}$. Inset figure shows the variation of $Nu/Ra_w^{1/3}$ with $Ra_w$; {\color{blue}---}, $\sim Ra_w^{-0.0059}$ and {\color{red}$---$}, $\sim Ra_w^{-0.0016}$. (b) Variation of mean local dimensionless shear with $Ra_w$; {\color{blue}---}, $0.734 Ra_w^{-0.102}$ and {\color{red}$---$}, $0.428 Ra_w^{-0.089}$. Top-right inset shows the variation of $\overline{u|}_{z=\delta}$ with $Ra_w$; {\color{blue}---}, $\sim Ra_w^{1/3}$ and {\color{red}$---$}, $\sim Ra_w^{0.38}$. Bottom-left inset shows the variation of $\overline{u|}_{z=\delta}/V_{bl}$ with $Ra_w$. {\color{blue}---}, $0.58 Ra_w^{-0.0067}$ and {\color{red}$---$}, $0.039 Ra_w^{0.059}$.}
\label{fig:avegedNu_A}
\end{figure}
\subsection{Average shear on the BLs and flux scaling}
\label{sec:flux_scals}
    We now estimate the dimensionless average thermal boundary layer thickness ($\overline{\delta_T}/H$) at each $Ra_w$ by numerically averaging the various values of  $\delta_T/H$ over the horizontal dimensionless plume spacing in the presence of shear ($\lambda_s/(2H)$), as 
    \begin{equation}
    \label{eq:deltH}
        \frac{\overline{\delta_T}}{H}=\frac{2H}{\lambda_s}\int_0^{\lambda_s/(2H)} \frac{\delta_T}{H}dx, 
    \end{equation}
  where, $\lambda_s$ is given by \eqref{eq:lam_lam0}. Using \eqref{eq:n_exponent}, for $Pr=1$, \eqref{eq:deltH} can be rewritten as,
  \begin{equation}
    \label{eq:deltH2}
        \frac{\overline{\delta_T}}{H}=\frac{2H}{\lambda_s}\int_0^{\lambda_s/(2H)} \frac{1}{C_2}\frac{\delta}{H}dx.
    \end{equation}
  We evaluate \eqref{eq:deltH2} using the $\delta$ obtained from the numerical solution of \eqref{eq:del_H} and \eqref{eq:A}, and $\lambda_s$ from \eqref{eq:lam_lam0}, with $Re$ obtained from \eqref{eq:cioni_Re} and \eqref{eq:Roche_Re} for the corresponding range of $Ra_w$, with the value of $C_2=0.72$, as suggested by \cite{shevkar_mohanan_puthenveettil_2023}. Using these values of $\overline{\delta_T}/H$ obtained from \eqref{eq:deltH2}, for each of the Reynolds number relations \eqref{eq:cioni_Re} and \eqref{eq:Roche_Re}, for the different $Ra_w$ corresponding to the range of \eqref{eq:cioni_Re} and \eqref{eq:Roche_Re}, we then obtain $Nu=H/(2\overline{\delta_T})$, for $\Gamma=1$ and $\Gamma=0.5$, respectively; the obtained variations of $Nu$ with $Ra_w$ are shown in figure~\ref{fig:Nu}. The inset in figure~\ref{fig:Nu} shows the corresponding variation of $Nu/Ra_w^{1/3}$ with $Ra_w$; $Nu/Ra_w^{1/3}\sim Ra_w^{-0.0059}$ for $\Gamma=1$ and $Nu/Ra_w^{1/3}\sim Ra_w^{-0.0016}$ for $\Gamma=0.5$. Figure~\ref{fig:Nu} shows the resulting $Nu$ vs $Ra_w$ scalings, namely, $Nu\sim Ra_w^{0.327}$ for $8\times10^{7}\leq Ra_w \leq 1\times10^{12}$ at $\Gamma=1$, and $Nu\sim Ra_w^{0.332}$ for $1\times10^{11}\leq Ra_w\leq 5\times10^{14}$ at $\Gamma=0.5$. Then, the present model where the local BLs in between the plumes, forced externally by the shear due to the LSF, determine the flux, predicts the classical 1/3 scaling of flux for the whole range of $8\times10^7\leq Ra_w\leq5\times10^{14}$; no transition to any ultimate regime is observed at $Ra_w=2.5\times10^{14}$ as suggested by~\cite{he2012_ultimateregm}.
  
  As we saw earlier, if the local BLs are purely natural convection  type, with the whole of $\Delta T_w$ occurring across them, when $Nu\sim Ra_w^{1/3}$, the classical scaling we observe for $Ra_w>10^{11}$. At the same time, in contrast, we showed in \S~\ref{sec:LBL and dim shear} that the shear at the edge of these local BLs, $u|_{z=\delta}$, increase with increase in $Ra_w$ (see the inset in figure~\ref{fig:Ax_Raw}), thereby making these local BLs deviate more from NCBLs. To understand this seemingly contradictory prediction of an increasing shear forcing at the edge of these BLs with increase in $Ra_w$, coinciding with a progressive tending towards a classical $1/3$ flux scaling, for $8\times10^{7}\leq Ra_w\leq5\times 10^{14}$, we now look at the average external shear on these local BLs, as well as the relative magnitude of the average shear forcing with respect to NCBL velocities, predicted by the present model. 

  We estimate the averaged dimensionless shear $\overline{A}$ at the edge of the local BLs by longitudinally averaging $A(x^{*})$, obtained by numerically solving \eqref{eq:del_H} and \eqref{eq:A}, over the distance $\lambda_s/(2H)$, using a similar expression as \eqref{eq:deltH2}, where $\delta/C_2H$ is replaced with $A(x)$. Figure~\ref{fig:A_Raw} shows the variation of $\overline{A}$ with $Ra_w$. $\overline{A}$ decreases with increase in $Ra_w$, similar to the reduction of $A(x^{*})$ with $Ra_w$ at any $x^*$, noticed earlier in figures \ref{fig:low_Ax}, \ref{fig:high_Ax} and \ref{fig:Ax_Raw}. Figure~\ref{fig:A_Raw} shows that $\overline{A}\sim Ra_w^{-0.102}$ for $Ra_w \leq 10^{12}$ at $\Gamma=1$ and $\overline{A}\sim Ra_w^{-0.089}$ for $Ra_w \geq 1\times 10^{11}$ at $\Gamma=0.5$, the exponents being almost the same as those obtained for $A(x)$. 
As in the case of $A(x^{*})$, $\overline{A}=\overline{u|}_{z=\delta
}/V_F$, with $V_F\sim (\nu/H)Ra_w^{3/7}$ for $Ra_w\leq10^{12}$ at $\Gamma=1$ and as $ (\nu/H)Ra_w^{0.48}$, for $Ra_w \geq1\times10^{11}$ at $\Gamma=0.5$, the average shear at the edge of the BLs, $\overline{u|}_{z=\delta}\sim (\nu/H)Ra_w^{1/3}$ for $Ra_w\leq10^{12}$ at $\Gamma=1$ and as $(\nu/H)Ra_w^{0.38}$ for $Ra_w\geq10^{11}$ at $\Gamma=0.5$, as shown in the top right inset in figure~\ref{fig:A_Raw}. Hence, the spatially averaged shear at the edge of the local BLs ($\overline{u}|_{z=\delta}$) increase with $Ra_w$, eventhough the dimensionless average shear at the edge of the BLs ($\overline{A}$) decreases with increase in $Ra_w$. Eventhough the shear forcing of the local BLs increases faster with $Ra_w$ at higher $Ra_w$ (as $Ra_w^{0.38}$, see top-right figure \ref{fig:A_Raw}) compared to that at lower $Ra_w$ (as $Ra_w^{1/3}$, see top-right figure \ref{fig:A_Raw}), the magnitude of shear at the higher $Ra_w$ are smaller, clearly due to the lower value of $\Gamma=0.5$ at the higher values of $Ra_w$. 

 The bottom left inset of figure~\ref{fig:A_Raw} shows the variation of $\overline{u|}_{z=\delta}/V_{bl}$ with $Ra_w$, where $V_{bl}$~\eqref{eq:Rebl} is the characteristic natural convection boundary layer velocity \citep{puthenveettil05:_plume_rayleig}. For $Ra_w \leq 10^{12}$ at $\Gamma=1$, $\overline{u|}_{z=\delta}/V_{bl}$ decreases with increase in $Ra_w$ as $\sim Ra_w^{-0.0067}$, while it increases with increase in $Ra_w$ for $Ra_w\geq 1\times 10^{11}$ at $\Gamma=0.5$, as $\sim Ra_w^{0.059}$. This changeover in scaling of $\overline{u|}_{z=\delta}/V_{bl}$ for $\Gamma=0.5$ occurs due to the change in dependence of $Re$ on $Ra_w$, from $Re\sim Ra_w^{3/7}$~\eqref{eq:cioni_Re} for $\Gamma=1$, to $Re\sim Ra_w^{0.48}$~\eqref{eq:Roche_Re} for $\Gamma=0.5$.  Similar changeovers in the scaling of $A(x^{*})$, $\overline{A}$ and $Nu$, due to this change in dependence of $Re$ on $Ra_w$, are observed for $\Gamma=0.5$ in figures~\ref{fig:Ax_Raw}, \ref{fig:A_Raw} and \ref{fig:Nu}, respectively.
 Eventhough $\overline{u|}_{z=\delta}/V_{bl}$ increase with increasing $Ra_w$ at $\Gamma=0.5$ for $1\times10^{11}\leq Ra_w\leq5\times10^{14}$, we note that it is always less than 0.3, and due to the lower $\Gamma$, much lesser than the corresponding values of $\overline{u}|_{z=\delta}/V_{bl}\simeq0.45$ at $\Gamma=1$. Hence, eventhough the average shear forcing ($\overline{u|}_{z=\delta}$) increases with increase in $Ra_w$ (see top-right inset in figure~\ref{fig:A_Raw}) natural convection velocities are still the dominant velocities within these local BLs, till $Ra_w=5\times10^{14}$, for both the aspect ratios, $\Gamma=1$ and $\Gamma=0.5$. 
 
 This dominance of the NC velocities within the local BLs, forced externally by the shear due to the LSF, even at $Ra_w$ as high as $5\times10^{14}$, explains the observation of the classical $1/3$ scaling of flux by the present model for $8\times10^7 \leq Ra_w \leq 5\times10^{14}$, with no transition to the ultimate regime seen at $Ra_w\geq 10^{13}$. Hence, eventhough the shear forcing of these BLs increase with increase in $Ra_w$ (see top right inset in figure \ref{fig:A_Raw}), the natural convection velocities dominate in these local BLs to give a classical 1/3 scaling of flux. An extrapolation of the trend of variation of $u|_{z=\delta}/V_{bl}$ for $\Gamma=0.5$ seen in the bottom inset of figure \ref{fig:A_Raw} to higher $Ra_w$ shows that $u|_{z=\delta}/V_{bl}\sim1$ at $Ra_w\approx10^{23}$; for higher $Ra_w$ we could expect the local BLs to become shear dominant, possibly resulting in a transition in flux scaling.  However this expectation must be qualified by the observation that the plume spacing, which determines the longitudinal extent of these local BLs, given by \eqref{eq:lam_lam0} is valid only till $Ra_w=5\times10^{16}$ at $\Gamma=0.5$. In any case \eqref{eq:lam_lam0} is a decreasing function of $Ra_w$, and hence it is possible that at some high $Ra_w$, the local BL length could become of the order of plume thickness; a new regime of flux scaling could be expected beyond such an $Ra_w$.

 \section{Conclusions}
Our study's key contribution lies in obtaining Nusselt number variation across a wide range of $8\times10^7\leq Ra_w\leq 5\times10^{14}$ through a numerical solution of a fifth-order algebraic scaling equation for boundary layers forced by the LSF in RBC. We propose a boundary layer model (refer figure~\ref{fig:local_bl}) which assumes the laminar natural convection BLs forced by the mean Prandtl-Blasius velocity profile generated due to the LSF velocity $V_F$. By integrating the BL equations (\S \ref{sec:integraleqs}) and using the appropriate scaling arguments for BL flow (\S \ref{sec:scaling rels}) and the shear forcing boundary condition at the edge of the BL \eqref{eq:von_karmann} in RBC, we obtained an algebraic equation (\ref{eq:del_H}). We use the expressions of Reynolds number based on the LSF velocity $V_F$ given by \cite{cioni} for $\Gamma=1$ and $Ra_w\le10^{12}$~\eqref{eq:cioni_Re} and \cite{Roche_2010} for $\Gamma=0.5$ and $1\times10^{11}\le Ra_w\le5\times10^{14}$~\eqref{eq:Roche_Re}. The algebraic equation was solved numerically to obtain the variation of thermal BL thickness with the longitudinal distance at various $Ra_w$. At any $Ra_w$, the averaged thermal BL thickness was then obtained by integrating $\delta_T$ over half the mean plume spacing $\lambda_s$ \eqref{eq:deltH2} and later this averaged thickness was used to obtain $Nu$. The results reveal that $Nu$ exhibits a power-law dependence on $Ra_w$, with an exponent of $m=0.327$ for $Ra_w\leq10^{12}$, while $m$ approaches $1/3$ for $1\times 10^{11}\leq Ra_w\leq 5\times10^{14}$. 

It has been shown that the dimensionless shear at the edge of the local boundary layer decreases as $Ra_w$ increases. The relationship between the average dimensionless shear and $Ra_w$ is given by $\overline{A}\sim Ra_w^{-0.102}$ for $8\times 10^7\leq Ra_w \leq 1\times10^{12}$ at $\Gamma=1$ and $\overline{A}\sim Ra_w^{-0.089}$ for $1\times 10^{11}\leq Ra_w \leq 5\times10^{14}$ at $\Gamma=0.5$. The average shear velocity acting at the edge of the boundary layer, denoted by $\overline{u|}_{z=\delta}$, is related to the dimensionless velocity scale $\overline{A}$ through the expression $\overline{A}=\overline{u|}_{z=\delta}/V_F$. For $Ra_w\le10^{12}$, the velocity scale $V_F$ scales as $Ra_w^{3/7}$, while for $Ra_w\ge1\times10^{11}$, $V_F$ scales as $Ra_w^{0.48}$. Thus, $\overline{u|}_{z=\delta}$ increases with increasing $Ra_w$,  scaling as $\overline{u|}_{z=\delta} \sim Ra_w^{1/3}$ for $8\times 10^7\leq Ra_w \leq 10^{12}$ at $\Gamma=1$, and $\overline{u|}_{z=\delta} \sim Ra_w^{0.38}$ for $1\times 10^{11}\leq Ra_w \leq5\times10^{14}$ at $\Gamma=0.5$. 
 Similarly, the local shear at the edge of the boundary layer, ${u|}_{z=\delta}$, increases with $Ra_w$, scaling as ${u|}_{z=\delta}\sim Ra_w^{0.331}$ for $8\times 10^7\leq Ra_w \leq 10^{12}$ at $\Gamma=1$, and ${u|}_{z=\delta}\sim Ra_w^{0.398}$ for $1\times 10^{11}\leq Ra_w \leq 5\times10^{14}$ at $\Gamma=0.5$. Eventhough, $\overline{u|}_{z=\delta}$ increased with increasing $Ra_w$, when normalised with the NCBL velocity $V_{bl}$, $\overline{u|}_{z=\delta}/V_{bl}$ decreases with $Ra_w$ for $Ra_w \leq 10^{12}$ and then exhibits a turnaround at $Ra_w \approx 10^{13}$, as shown in the inset of figure~\ref{fig:A_Raw}. However, this turnaround in the scaling of the shear velocity acting at the edge of BLs does not induce a transition in the flux scaling to the so called ultimate regime for $Ra_w\leq10^{15}$. The most probable reason is the much smaller than one values of $\overline{u|}_{z=\delta}/V_{bl}$, the sub-dominant shear forcing acting on the BLs compared to the NCBL velocities within the BLs for the given range of $Ra_w$.

Based on the aforementioned theoretical findings for $Ra_w\le10^{15}$, it is suggested that the classical scaling of flux was observed due to the sub-dominance of shear forcing on boundary layers relative to natural convection velocities within the boundary layers. Our observations does not indicate any possibility of the ultimate regime until $Ra_w=10^{15}$. This is in contrast to the suggestion made by He et al. (2012) and consistent with the results of Iyer et al. (2019) and \cite{lindborg_2023}. For $Ra_w>10^{15}$, the shear effects would be dominant when $\overline{u|}_{z=\delta}/V_{bl}\gg1$. If the trend of $\overline{u|}_{z=\delta}/V_{bl}$ with $Ra_w$ at $\Gamma=0.5$, shown in the inset of figure~\ref{fig:A_Raw}, continues, it suggests that values of $\overline{u|}_{z=\delta}/V_{bl}\gg1$ could potentially cause BLs to become turbulent, leading to a transition in the flux scaling. As per the trend at $\Gamma=0.5$, this transition would then occur at $Ra_w\approx 10^{23}$ , in the range of $Ra_w$ proposed originally by \cite{Kraichnan1962}, i.e., $10^{21} \leq Ra_w\leq 10^{24}$. Furthermore, the transition in the flux scaling for $\Gamma\sim1$ and $Pr\sim1$ could occur at $Ra_w$ slightly less than $10^{23}$. 

In conclusion, our analysis shows that laminar natural convection-type boundary layers (BLs), when subjected to a sufficiently large driving strength similar to that in Rayleigh-Bénard convection (RBC), are not expected to remain laminar for $Ra_w\gg 10^{15}$. They may first become Prandtl-Blasius type and then transition to a turbulent state, leading to a change in flux scaling. Finally, the LSF cannot be defined solely by scaling laws but must be treated as a statistical quantity. In addition, some regions of boundary layers get strongly affected by the shear due to the dynamically varying LSF and others are not~\citep{shevkar_mohanan_puthenveettil_2023}. The dynamic, wide-ranging LSF observed close to the plate highlights the critical need to study boundary layers at the local scale, as opposed to the global scale in Rayleigh B\'enard convection. Further investigations for $Ra_w\ge10^{15}$, utilizing high-resolution DNS and experimental data, are necessary to study the effect of shear generated by the LSF on local boundary layers at various spatial locations. Such studies are crucial to clarify the complete picture, including the sub-critical transitions.

 \nocite{*}
\bibliography{refs}
 \bibliographystyle{jfm}

\end{document}